\documentclass[aps,prl,reprint,superscriptaddress,twocolumn]{revtex4-2}
\usepackage{graphicx}
\usepackage{amsfonts}
\usepackage{amsthm}
\usepackage{amsmath}
\usepackage{amsfonts}
\usepackage{amssymb}
\usepackage{centernot}
\usepackage{bm}
\usepackage{braket}
\usepackage[normalem]{ulem}
\usepackage[colorlinks=True,linkcolor=red,citecolor=blue,urlcolor=blue]{hyperref}
\usepackage[dvipsname]{xcolor}

\usepackage{pdfpages}
\makeatletter
\AtBeginDocument{\let\LS@rot\@undefined}
\makeatother

\usepackage{times}
\newcommand\eq[1]{\begin{align}#1\end{align}}

\newcommand{\am}{\alpha_{m}}
\newcommand{\at}{\alpha_{t}}

\newcommand{\comment}[1]{}

\begin{document}
\title{Anomalous multifractality in quantum chains with strongly correlated disorder}

\author{Alexander Duthie}
\email{alexander.duthie@chem.ox.ac.uk}
\affiliation{Physical and Theoretical Chemistry, Oxford University,
South Parks Road, Oxford OX1 3QZ, United Kingdom}

\author{Sthitadhi Roy}
\email{sthitadhi.roy@chem.ox.ac.uk}
\affiliation{Physical and Theoretical Chemistry, Oxford University,
South Parks Road, Oxford OX1 3QZ, United Kingdom}
\affiliation{Rudolf Peierls Centre for Theoretical Physics, Clarendon Laboratory, Oxford University, Parks Road, Oxford OX1 3PU, United Kingdom}

\author{David E. Logan}
\email{david.logan@chem.ox.ac.uk}
\affiliation{Physical and Theoretical Chemistry, Oxford University,
South Parks Road, Oxford OX1 3QZ, United Kingdom}
\affiliation{Department of Physics, Indian Institute of Science, Bangalore 560012, India}

\begin{abstract}
We demonstrate numerically that a robust and unusual multifractal regime can emerge in a one-dimensional quantum chain with maximally correlated disorder, above a threshold disorder strength. This regime is preceded by a mixed and an extended regime at weaker disorder strengths, with the former hosting both extended and multifractal eigenstates. The multifractal states we find are markedly different from conventional multifractal states in their structure, as they reside approximately uniformly over a continuous segment of the chain, and the lengths of these segments scale non-trivially with system size. This anomalous nature also leaves imprints on dynamics. An initially localised wavepacket shows ballistic transport, in contrast to the slow, generally subdiffusive, transport 
commonly associated with multifractality. However, the timescale over which this ballistic transport persists again scales non-trivially with the system size.
\end{abstract}

\maketitle

Multifractal wavefunctions in quantum systems, which are neither extended nor localised, are characterised by anomalous statistics of their amplitudes~\cite{evers2008anderson}. While the effective volume occupied by such states grows unboundedly with system size, it is a vanishing fraction of the system volume; as such, they are often 
dubbed non-ergodic extended states. This is reflected in the scaling of the moments of the wavefunction amplitudes with system size. For a wavefunction $\psi(x)$ defined on a discrete graph with $L$ sites,
\eq{
    \sum_{x=1}^L|\psi(x)|^{2q}\sim
    \begin{cases}
    L^{-(q-1)} & :\mathrm{extended}\\
    L^{-D_q(q-1)} &
		:\mathrm{multifractal}\\
    L^0 &  :\mathrm{localised}
    \end{cases}\,
    \nonumber
} 
where $0<D_q<1$ is the so-called multifractal dimension~\cite{evers2008anderson}.

In the context of short-ranged disordered systems in finite-dimensions, multifractality is often a feature of critical points, such as Anderson transitions~\cite{evers2008anderson,wegner1980inverse,castellani1986multifractal,chalker1988scaling,chalker1990scaling,bauer1990correlation,schreiber1991multifractal,janssen1994multifractal,chalker1996spectral,mirlin2000multifractality,evers2000fluc,cuevas2001fluctuations}
and quantum Hall plateau transitions~\cite{chalker1988percolation,huckenstein1995scaling,evers2001multifractality}, which are clearly fine-tuned points in parameter space. Multifractality is also realised, often  robustly, in several long-ranged disordered hopping models, and fully connected random-matrix ensembles~\cite{levitov1990delocalisation,mirlin1996transition,parshin1998multifractal,levitov1999critical,varga2000critical,cuevas2001anomalously,cuevas2003multifractality,monthus2010anderson,kravtsov2015random,deng2019one,monthus2019multifractality,khaymovich2020fragile,kravtsov2020localization,biroli2021levy}. The presence of long-ranged physics, either emergently via diverging correlation lengths in the former, or explicitly via the structure of the models in the latter, unifies the two contexts. An interesting question thus arises: how can a robust multifractal phase be realised in a quantum system with inherently short-ranged interactions/hoppings?  One possible avenue in a manifestly out-of-equilibrium setting is the time-periodic modulation of a quasiperiodic system with a mobility edge or a localisation transition~\cite{roy2018multifractality,sarkar2021mobility}.

In this work, we demonstrate an alternative pathway to robust multifractality in an inherently short-ranged system, importantly in a time-independent Hamiltonian setting. The central ingredient is strong (in fact \emph{maximal}, as 
clarified below) correlations in the disordered on-site potential of a one-dimensional chain. Interestingly, the origin and  resultant structure of the multifractal states in such a system is markedly different from that of conventional multifractal states, which are associated with both rare large peaks and long polynomial tails of wavefunction amplitudes. In contrast, the multifractal states we find in this work reside over continuous
segments of length $\ell$ in the chain and crucially, within the segments, the wavefunction intensities are approximately \emph{uniform}; $\vert\psi(x)\vert^2\approx 1/\ell$, and 0 elsewhere.  Due to this structure, we refer to the states as \emph{tabletop} states. The multifractality of the wavefunctions is then encoded in the scaling of these tabletop lengths with system size. The anomalous nature of the multifractal states also has implications for dynamics. Multifractality of eigenstates is often accompanied by slow dynamics~\cite{detomasi2016generalised,monthus2017multifractality,roy2018multifractality,khyamovich2021dynamical},\footnote{It should be noted that multifractality is not always a prerequisite for slow dynamics~\cite{detomasi2020subdiffusion,khyamovich2021dynamical}.}. However, in this case, we find ballistic spreading of a initially localised wavepacket, but over timescales that scale non-trivially with the system size.

\begin{figure}[!t]
\includegraphics[width=\linewidth]{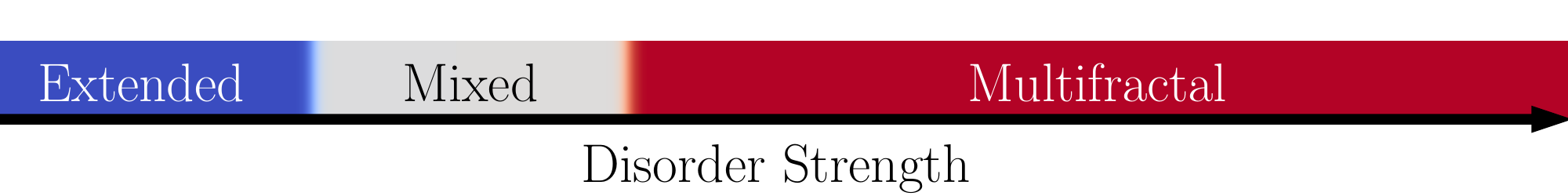}\caption{
Schematic of the three regimes of the model Eq.~\ref{eq:H},  as a function of disorder strength. For small disorder,
all states are extended, and for large enough disorder all states are multifractal. In between these two phases is a mixed regime, where the disorder-averaged spectrum hosts a mixture of extended and multifractal states. }\label{fig:schematic}
\end{figure}

\begin{figure}[!t]
\includegraphics[width=\linewidth]{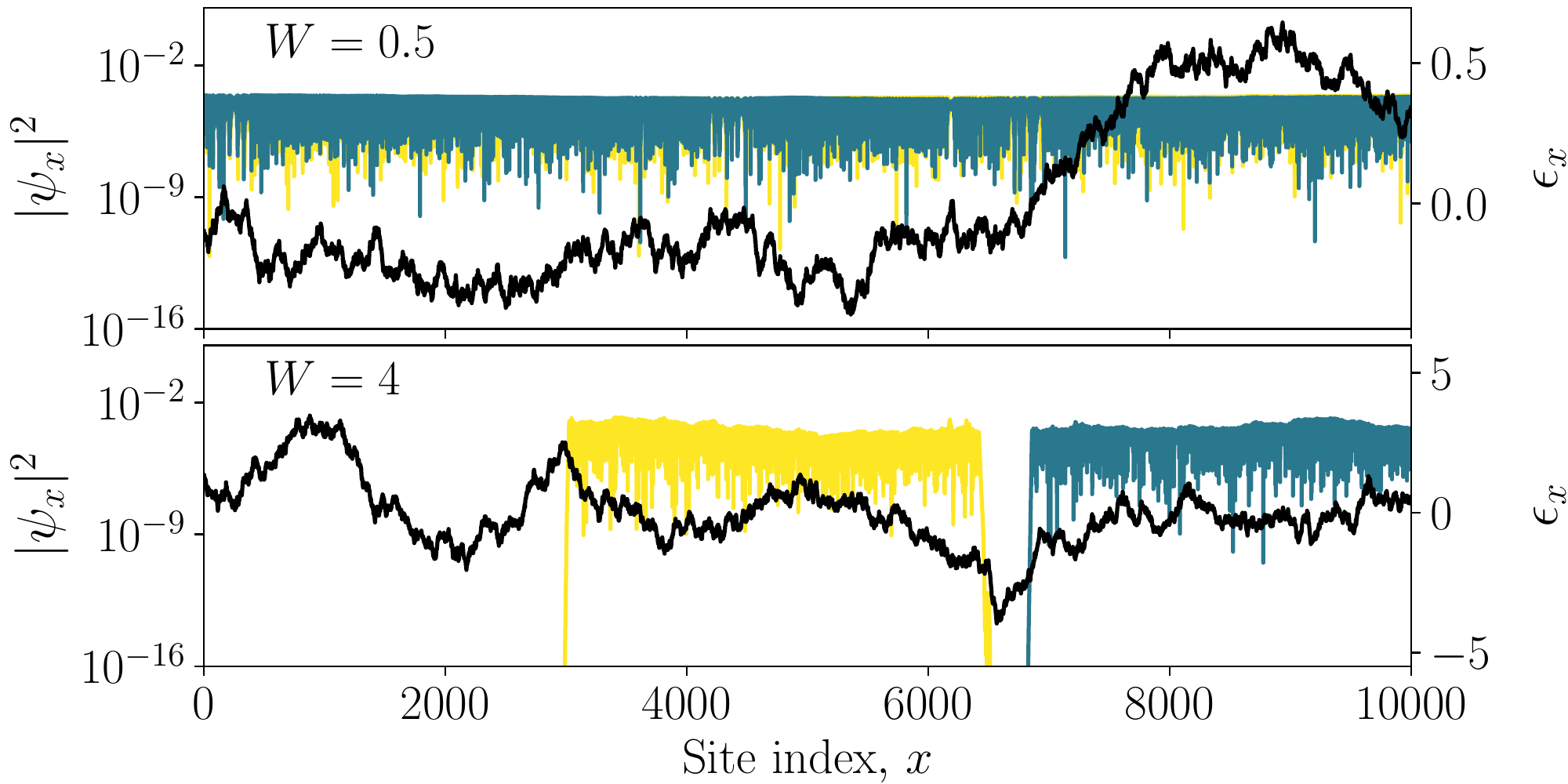}
\caption{For weak and strong disorder strengths $W$ (top and bottom respectively), representative eigenstates are shown in blue and yellow. In the former, tabletops span the entire chain, whereas in the latter they span a sub-extensive segment. Black traces denote a realisation of the maximally correlated disorder potential (labels on right axis) wherein short distance fluctuations are visibly suppressed.}
\label{fig:eigs}
\end{figure}

As a concrete model, we consider a disordered tight-binding Hamiltonian on a chain of length $L$,
\begin{equation}
H = W\sum_{x=1}^L\epsilon_x|x\rangle\langle x| + \sum_{x=1}^{L-1}\left[|x\rangle\langle x+1| + \ket{x+1}\bra{x}\right]\,,\label{eq:H}
\end{equation}
where $W$ denotes the disorder strength and the on-site potentials are drawn from a multivariate Gaussian distribution, $\vec{\epsilon_x}\sim\mathcal{N}(\mathbf{0},\mathbf{C})$, with zero mean. The correlations in the potential are encoded in the covariance matrix $\mathbf{C}$. We take the correlations in the potential to decay with distance as
\eq{
    C(r)\equiv[\mathbf{C}]_{x,x+r}=\braket{\epsilon_x\epsilon_{x+r}}=f(r/L);\quad f(0)=1.
    \label{eq:corr}
}
The key point here is that the correlation is a function of $r/L$, which implies that in the thermodynamic limit $C(r)\to 1$ for all sub-extensive $r$, as $\lim_{L\to\infty}(r/L)\to 0$. In other words, the potentials of two sites a sub-extensive distance $r$ apart are completely  slaved to each other in the thermodynamic limit. We refer to this as \emph{maximal correlations} in the disorder~\cite{roy2020correlations,roy2020strongly}. For specificity 
we choose $f(r/L)=\exp[-r/(\lambda L)]$, but emphasise that the specific functional form of $f$ is immaterial. In the following we set $\lambda=1$. Note that the limit of $\lambda\to0$ is singular, as there the model becomes the conventional $1$D Anderson model with all eigenstates exponentially localised~\cite{anderson1958absence,mott1961theory}.

The form of the correlations, \eqref{eq:corr}, endows the disordered potential with an extensive lengthscale ($\lambda L$ in this case), such that the potential fluctuations on sub-extensive scales are heavily suppressed and only those at extensive scales survive for large $L$. This is already suggestive that the eigenstates can be extended over lengthscales which scale non-trivially with $L$, resulting in multifractality. In fact, as demonstrated below, we find three distinct regimes as a function of $W$. For sufficiently strong disorder a robust multifractal phase is found, where the average or typical tabletop lengths scale as $L^\alpha$ with $\alpha<1$; whereas for weak disorder we find $\alpha=1$  for all eigenstates, indicating an extended phase. For a range of $W$ between these two regimes a mixed phase is found, where for a given energy some realisations host extended states and some multifractal. Establishing these robust multifractal and extended regimes with the intervening mixed regime in a model with maximally correlated disorder is the central result of this work, and is summarised in Fig.~\ref{fig:schematic}.

Before delving into a detailed analysis of the statistics of eigenstate tabletop lengths and the consequent multifractality, in Fig.~\ref{fig:eigs} we show explicitly the tabletop nature of eigenstates. Operationally, we extract the tabletop edges for an eigenstate by scanning the chain for sites where the $\vert\psi_x\vert^2$ jumps from zero (within numerical precision) and the tabletop length $\ell$ is then simply the distance between the two such sites. It is evident from Fig.~\ref{fig:eigs} that the eigenstates are approximately uniform over the tabletop segments. Hence it is natural to study the distribution of the tabletop lengths $\ell$, or equivalently of $\tilde{\ell}=\ell/L$; denoting these distributions by $P_\ell$ and $P_{\tilde{\ell}}$ respectively. Since we are interested
in particular how $\ell$ scales with system size $L$, we define the exponents $\am$ and $\at$ from the mean and typical tabletop lengths as
\eq{
    \ell_m\equiv\braket{\ell}\sim L^{\am};\quad \ell_t\equiv\exp[\braket{\log \ell}]\sim L^{\at}\,,
    \label{eq:alphamt}
}
where $\braket{\ell}=\int d\ell~\ell P_\ell(\ell)$ and similarly for $\braket{\log \ell}$.
In addition, we also define the exponent $\alpha=\log\ell/\log L$, and study its distribution which we denote by $P_\alpha$. 

\begin{figure}
\includegraphics[width=\linewidth]{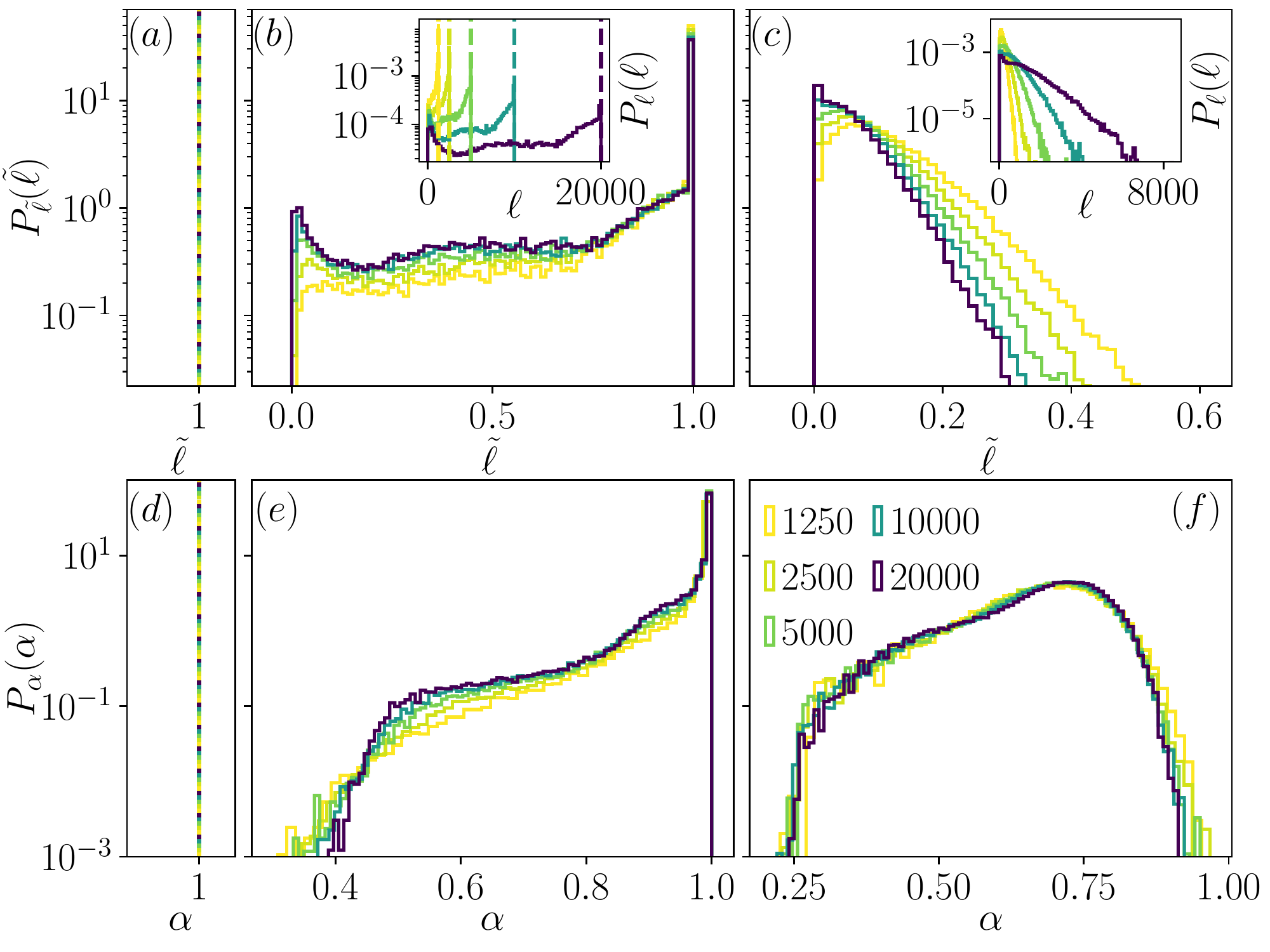}
\caption{(a)-(c): Main panels show distributions $P_{\tilde{\ell}}(\tilde{\ell})$ of $\tilde{\ell}=\ell/L$
for representative disorder strengths in the extended ($W=0.5$), mixed ($W=2$), and multifractal ($W=8$) regimes, for different system sizes $L$ (indicated in panel (f)). Insets show the corresponding $P_\ell(\ell)$ distributions.  (d)-(f): The distribution $P_\alpha(\alpha)$ for the corresponding $W$ values. Statistics are obtained over $20000$ disorder realisations.}
\label{fig:tt_dists}
\end{figure}

We turn now to numerical results, which unless stated otherwise refer to band centre states (results for other energies remain qualitatively the same). For weak disorder, we find that the tabletops span not only an extensive segment of the chain but the entirety of it, such that $P_\ell(\ell)=\delta(\ell-L)$ or equivalently $P_\alpha(\alpha)=\delta(\alpha-1)$ [Figs.~\ref{fig:tt_dists}(a) and \ref{fig:tt_dists}(d)]. This is the extended regime indicated
in Fig.~\ref{fig:schematic}. On increasing $W$, the distribution $P_\ell(\ell)$ develops finite weight at sub-extensive $\ell$ while retaining the rest of the weight at extensive $\ell$ [Figs.~\ref{fig:tt_dists}(b) and \ref{fig:tt_dists}(e)]. This is the mixed regime referred to in Fig.~\ref{fig:schematic}. Note that it is important to distinguish the multifractal states with sub-extensive $\ell$ from those that occupy an extensive $\ell=a L$ with 
$a\leq 1$, and hence confirm the presence of the former. That this is indeed the case is evidenced in Fig.~\ref{fig:tt_dists}(b) where the weight of the distribution $P_{\tilde{\ell}}(\tilde{\ell})$ at $\tilde{\ell}=0$ grows with increasing $L$. Further clear evidence for the mixed regime is also seen in Fig.~\ref{fig:tt_dists}(e) by the fact that $P_\alpha(\alpha)$ retains weight at both $\alpha=1$ and $\alpha<1$ as $L\to\infty$.  Finally, at strong disorder, the system enters a regime where all states in the spectrum are multifractal. As shown in Fig.~\ref{fig:tt_dists}(c), for  finite $\tilde{\ell}$ the weight in $P_{\tilde{\ell}}(\tilde{\ell})$ ultimately decays with increasing $L$, suggesting that there exist no states with extensive tabletop lengths. That all states are 
indeed  multifractal is further confirmed by the distribution $P_\alpha(\alpha)$ (which is well converged with $L$) having support strictly on $0<\alpha<1$ as shown in Fig.~\ref{fig:tt_dists}(f). The  vanishing of the weight of $P_\alpha(\alpha)$ at $\alpha=1$ and $\alpha=0$ implies respectively the absence of extended and localised states.
We add that this three-phase picture is also consistent with results for the scaling of transmittances with $L$, obtained via a transfer matrix calculation~\cite{supp}.

\begin{figure}[!t]
\includegraphics[width=\linewidth]{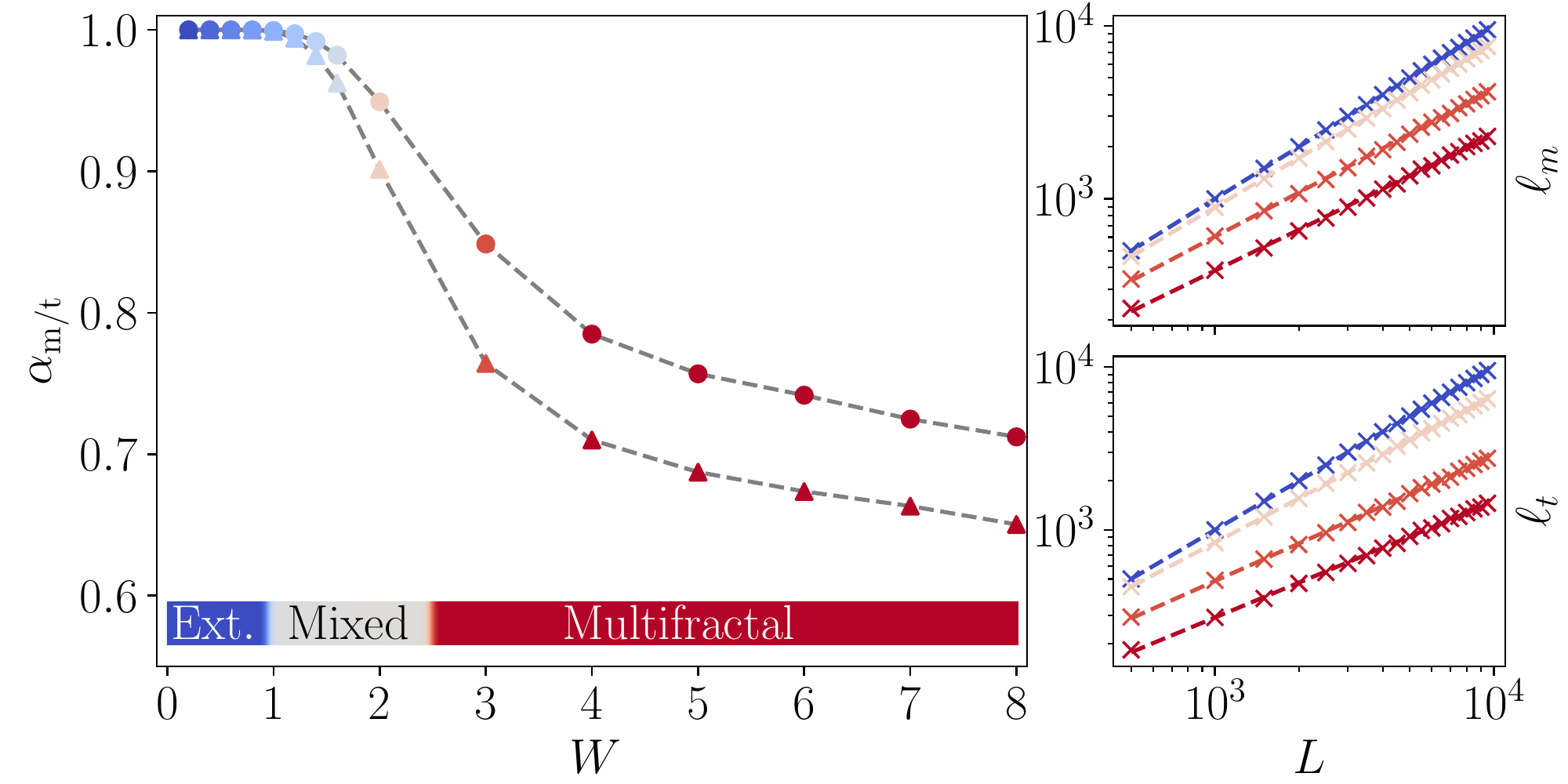}
\caption{Left: Exponents $\am$ (circles) and $\at$ (triangles), defined in Eq.~\ref{eq:alphamt}, as a function of disorder strength $W$. Right: Fits of $\ell_{m}$ and $\ell_t$ versus $L$ on logarithmic axes, used to extract the exponents. Data shown for $W=0.5$, $2$, $3$, and $4$.}
\label{fig:tt_scaling}
\end{figure}

Having established the presence of a multifractal regime, along with an extended and a mixed regime, based on the distributions $P_{\tilde{\ell}}$ and $P_{\alpha}$, we next present results for the scaling of mean and typical tabletop lengths with $L$. In particular, Fig.~\ref{fig:tt_scaling} shows results for $\am$ and $\at$ (defined in Eq.~\ref{eq:alphamt}) as a function of $W$. For weak disorder we find both $\am=1$ and $\at=1$, consistent with the presence solely of extended states. On increasing $W$ and entering the mixed regime, $\at$ and $\am$ decrease from 1, indicating the emergence of multifractal states in the spectrum. Note that on entering the mixed regime, $\am$ deviates from 1 less markedly than $\at$. This is natural, as in the presence of both extended and multifractal states
the mean $\ell_m$ is dominated by the extended states with $\ell\sim L$, whence $\am$ is closer to 1 than $\at$.
Finally, on increasing $W$ further into the regime where the spectrum has solely multifractal states, $\am$ and $\at<\am$ continue to decrease monotonically.

\begin{figure}[!b]
\includegraphics[width=\linewidth]{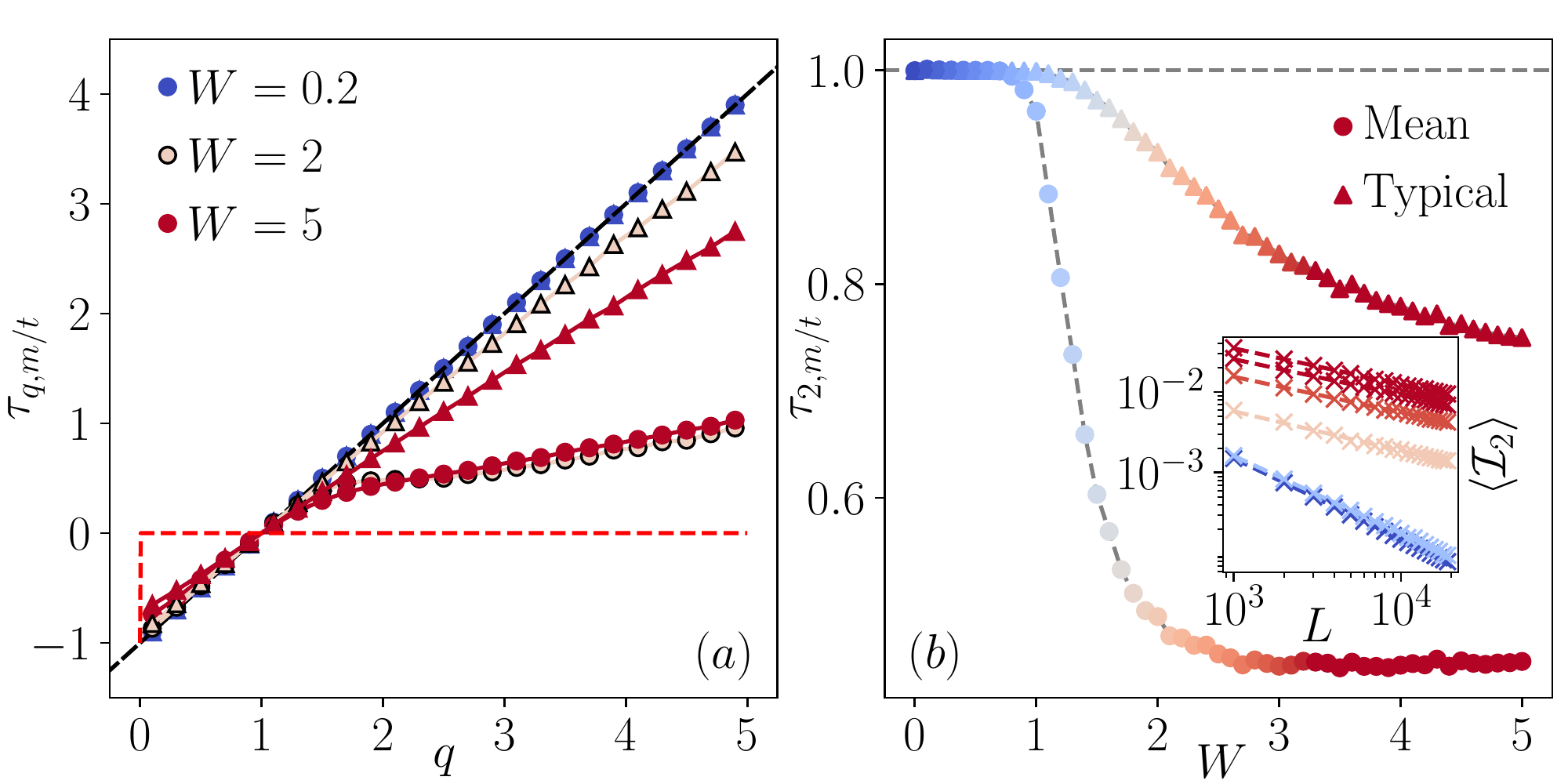}
\caption{(a) Exponents $\tau_{q,{m}}$ (circles) and $\tau_{q,t}$ (triangles), defined in Eq.~\ref{eq:tauqmt}, for representative $W$ in the three regimes. Black and red dashed lines respectively denote extended and localised behaviour. (b) Exponents $\tau_{2,{m}}$ and $\tau_{2,{t}}$ as a function of  $W$. Inset shows representative plots of $\braket{\mathcal{I}_2}$ versus $L$, used to extract $\tau_{2,m}$, for $W=0,1,2,3,4,5$.}
\label{fig:IPR}
\end{figure}

We close our analysis of the multifractal statistics of tabletop eigenstates with results of a standard probe of multifractality, the generalised inverse participation ratios (IPR), defined as $\mathcal{I}_q(\psi) = \sum_{x=1}^L|\psi(x)|^{2q}$. We will be interested in the scaling with $L$ of both the mean and typical IPR,
\eq{
    \mathcal{I}_{q,m}=\braket{\mathcal{I}_q}\sim L^{-\tau_{q,{m}}},~~\mathcal{I}_{q,t}=\exp[\braket{\log\mathcal{I}_q}]\sim L^{-\tau_{q,{t}}}\,.
    \label{eq:tauqmt}
}
Extended states have $\tau_{q,{m/t}}=q-1$, while for exponentially localised states $\tau_{q,{m/t}}=0$ for $q>0$. An intermediate behaviour for $\tau_q$ indicates multifractal states~\cite{evers2008anderson}. Fig.~\ref{fig:IPR}(a) shows results for $\tau_{q,{m}}$ and $\tau_{q,{t}}$, for representative $W$ values in each of the three regimes. For the weakest disorder, which lies in the extended regime, we indeed find $\tau_{q,{m/t}}=q-1$. The presence of multifractal states upon increasing $W$ is borne out by $0<\tau_{q,{m/t}}<q-1$ for $q>1$. In Fig.~\ref{fig:IPR}(b), we focus on $\tau_{2,m/t}$ as a function of $W$. Note that on increasing $W$ and entering the mixed regime, $\tau_{2,{m}}$ deviates from the ergodic value of 1 more markedly than $\tau_{2,t}$, reflecting the fact that the mean IPR is 
dominated by the small fraction of multifractal states which have qualitatively larger IPRs than the extended ones.

An explanatory comment is in order regarding the lack of energy resolution between extended and multifractal states in the mixed regime. The density of states (DoS) for the model \eqref{eq:H} with the correlated disorder \eqref{eq:corr} can be shown to fluctuate strongly across disorder realisations~\cite{supp}. For each realisation we choose to refer energies relative to the centre of the spectrum (tantamount to $H\to H-\mathrm{Tr}[H]$). However, the fluctuations in all higher moments of the DoS also remain finite in the thermodynamic limit. As a result, whether 
an eigenstate at a given energy (relative to the centre of the spectrum) is extended or multifractal depends on the specific disorder realisation (though for any given realisation, multifractal and extended states 
do not of course coexist  at the same energy). Averaging over disorder realisations therefore smears out any energy resolution in the mixed regime, thereby precluding the traditional notion of a mobility edge in the averaged DoS.

So far, we have focussed on `static' properties of the model. Since multifractality often goes hand in hand with slow dynamics~\cite{detomasi2016generalised,monthus2017multifractality,roy2018multifractality,khyamovich2021dynamical}, it is worth asking what imprint the anomalous multifractal states leave on the dynamics in the present case. In order to answer this question, we consider the spreading of an initially localised wavepacket, in particular its second moment defined as
\eq{
    X^2(t)=\Big\langle\sum_{x=1}^L x^2|\psi(x,t)|^2-\big[\sum_{x=1}^Lx|\psi(x,t)|^2\big]^2\Big\rangle\,,
    \label{eq:xsqt}
}
with $\psi(x,t=0)=\delta_{x,L/2}$. The simplest expected behaviour for $X^2(t)$ is 
\eq{
    X^2(t)\sim\begin{cases}
            t^{2z};\quad &t\ll t_\ast\\
            L^{2\beta};\quad &t\gg t_\ast
            \end{cases}\,,
}
with crossover scale $t_\ast\sim L^{\beta/z}$. For example, $z,\beta=1$ would correspond to ballistic transport
until the wavepacket spans the entire system; while for localised states $z,\beta=0$, indicating the absence of transport. Conventionally, multifractal states lead to $0<z,\beta<1$~\cite{detomasi2016generalised,monthus2017multifractality,roy2018multifractality,khyamovich2021dynamical}.

\begin{figure}
\includegraphics[width=\linewidth]{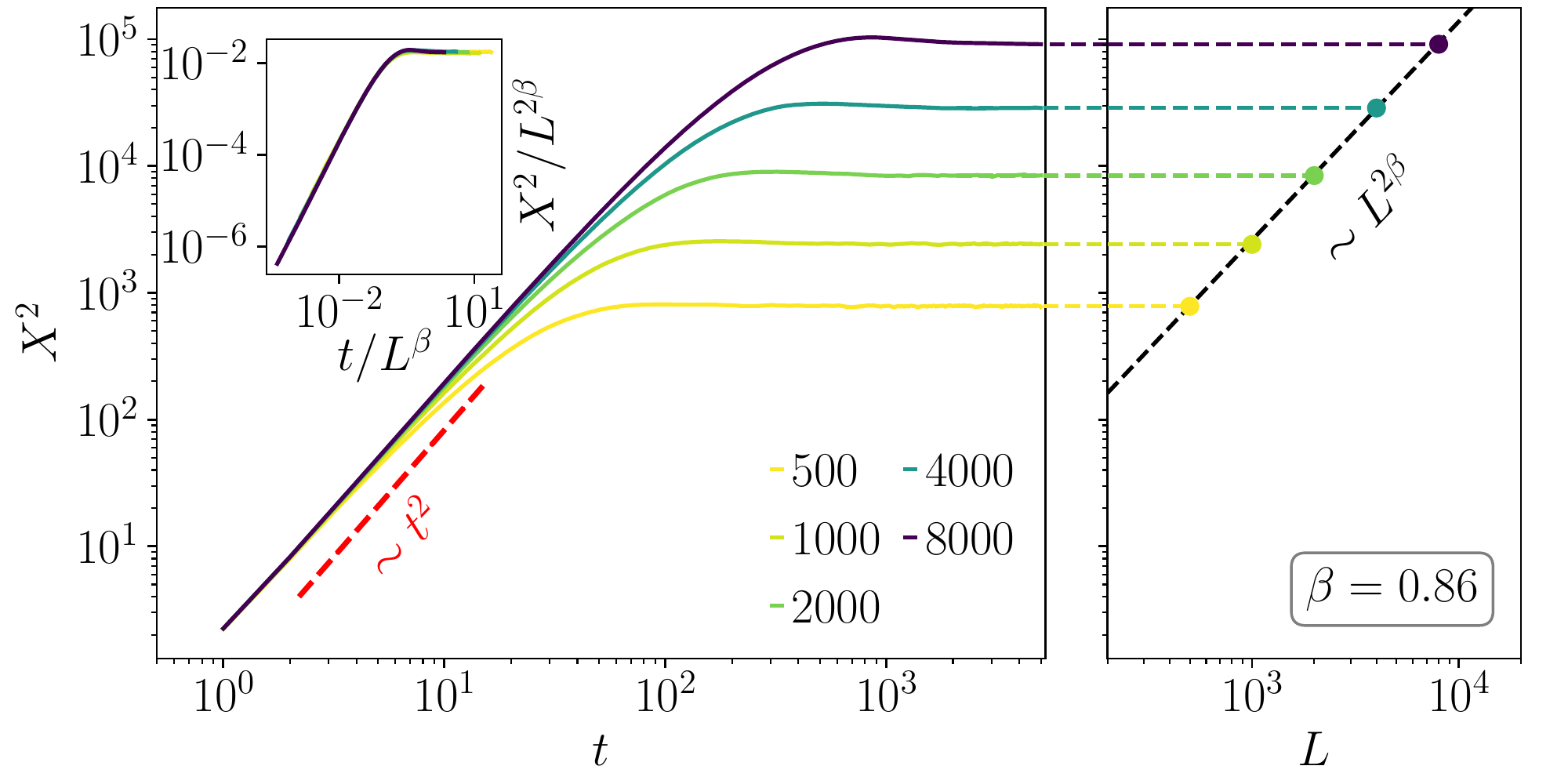}
\caption{Dynamics of an initially localised wavepacket in the multifractal regime, for $W=6$. Left: Ballistic transport of the wavepacket as $X^2(t)$ (defined in Eq.~\ref{eq:xsqt}) initially grows as $t^2$. Right: The saturation value, $X^2(t\to\infty)\sim L^{2\beta}$ plotted as a function of $L$, with $\beta \simeq 0.85$. The inset shows the scale-collapsed form $X^2(t)=L^{2\beta}g(t/L^\beta)$, with $g(y)\sim y^2$ for $y\ll 1$.}
\label{fig:dynamics}
\end{figure}

Fig.~\ref{fig:dynamics} shows results for the maximally correlated model in the pure multifractal phase. Remarkably, the transport is ballistic with $z=1$ but the saturation of $X^2(t)$ scales as $L^{2\beta}$ with exponent 
$\beta <1$. In fact, the inset in Fig.~\ref{fig:dynamics} shows the scale-collapsed behaviour $X^2(t) = L^{2\beta}g(t/L^\beta)$ with $g(y)\sim y^2$ for $y \ll 1$, which confirms the ballistic spreading and sub-extensive saturation of the wavepacket. We now show how a toy model of the anomalous multifractal states qualitatively rationalises the above dynamics.

For simplicity, let us assume that in a given realisation the initially localised wavepacket lies at the centre of a tabletop segment of length $\ell$. Approximating the eigenstates therein as fully extended over the segment, the wavepacket spreads as a truncated Gaussian
\eq{
    |\psi(x,t)|^2 = \Theta(\ell-2|x|)e^{-x^2/t^2}/\big(t\sqrt{\pi}~\mathrm{Erf}(\ell/2t)\big)\,,
    \label{eq:psixt}
}
where we have absorbed the velocity into the units of time. $X^2(t)$ for the wavepacket \eqref{eq:psixt} behaves in time as 
\eq{
    X_\ell^2(t) =\tfrac{1}{2} t^2[1-e^{-\ell^2/4t^2}(\ell/t)/\big(\sqrt{\pi}~\mathrm{Erf}(\ell/2 t)\big)]\,.
}
Averaging over disorder is equivalent to integrating over the probability distribution of the initial site residing in a tabletop of length $\ell$, denoted by $\Phi_\ell(\ell, L)$. Thus
\eq{
    X^2(t) = \tfrac{1}{2}t^2
    \left[1-\frac{2}{\sqrt{\pi}}\int d\ell~\Phi_\ell(\ell,L)\frac{\ell e^{-\ell^2/4t^2}}{2t~\mathrm{Erf}(\ell/2t)}\right]\,.
    \label{eq:xsqtoy}
}
For timescales much smaller than a typical tabletop length, the second term in Eq.~\ref{eq:xsqtoy} is negligibly small and the ballistic behaviour $X^2(t)\sim t^2$ is recovered. In the opposite limit where $t$ is much larger than typical tabletop lengths (such that $\ell/2t\ll 1$ in \eqref{eq:xsqtoy}), the wavepacket spreading saturates and $X^2\simeq \int d\ell~\Phi_\ell(\ell,L)\ell^2$, which one expects to scale non-trivially with $L$  due to that of $\Phi_\ell(\ell,L)$.

Finally, we provide a simple heuristic argument for the physics underlying the emergence of the multifractal regime. Note that with the form of the correlations, $C(r)=e^{-r/L}$, the sequence of site energies $\dots,\epsilon_i,\epsilon_{i+1},\dots$ is readily shown to be a martingale. Thus, given $\epsilon_i$, $\epsilon_{i+1}$ is a Gaussian random number with mean $e^{-1/L}\epsilon_i$ and variance $\simeq 2/L$. This leads to the site energies of nearby sites being very close to each other  with  high probability; to which we attribute the approximately uniform $|\psi(x)|^{2}$ across the tabletop segment. However, since the conditional distribution $P(\epsilon_{i+1}|\epsilon_i)$ is a Gaussian, it has unbounded support. As a result, with very low probability, there can be neighbouring sites where the site energies are wildly different. The break in the tabletops may be attributed to such rare regions. Moreover,
as the sequence of site energies is a martingale, the probability of a tabletop terminating at a site is  independent of the energies of the sites prior to it. It is then intuitively natural to regard the tabletop breaks 
as a Poisson process. This in turn  implies $P_\ell(\ell)\sim e^{-\gamma \ell}$ where $\gamma$ is the density of tabletop breaks, which decays with increasing $L$~\footnote{As each tabletop state has two breaks, and the density of tabletops of length $\ell\sim L^{\alpha}$ is $\sim L^{-\alpha}$.}. An exponential $P_\ell(\ell)$ with a rate that decays with $L$ is indeed consistent with the numerical results shown in Fig.~\ref{fig:tt_dists}(c) in the strong disorder regime. Obviously,  however, obtaining the precise multifractal exponents would require a theory which takes
into account the fluctuations of $|\psi(x)|^2$ within the tabletop length.

In summary, we demonstrated numerically that a one-dimensional quantum chain~\eqref{eq:H}, with maximally correlated disorder~\eqref{eq:corr}, hosts a robust multifractal regime. These multifractal eigenstates are strikingly unusual. They reside approximately uniformly over segments of the chain whose lengths scale non-trivially with $L$, whence the multifractal statistics arise. Such an anomalous structure also leaves imprints on the dynamics -- in the multifractal phase an initially localised wavepacket spreads ballistically, but over timescales that scale as $L^\beta$ with $\beta<1$, beyond which the expansion saturates. 

We note that the behaviour found here differs radically~\footnote{Perhaps expectedly, given that the Hausdorff dimension of any tree with $K\geq 2$ is infinite, while that of a 1D chain is unity.} from the same model considered on a tree with connectivity $K\geq 2$~\cite{roy2020strongly},  which hosts localised states but not a multifractal 
phase. The present work is also very different from earlier studies~\cite{izrailev1999localisation,croy2011anderson} 
of 1D Anderson localisation with long-ranged, power-law (and hence scale-free) disorder correlations. These models
too do not host a robust multifractal regime. This suggests that the system-size dependent scale introduced by the correlations \eqref{eq:corr}, along with the martingale nature of the site-energies, is responsible for the robust multifractal regime in the 1D model with maximally correlated disorder.

Finally, it is interesting to note that the effective Fock-space disorder in a many-body localised system is likewise
maximally correlated~\cite{roy2020correlations}, and the many-body localised eigenstates indeed exhibit multifractality on the Fock space~\cite{deluca2013ergodicity,mace2018multifractal,detomasi2021PRB,roy2021fockspace}. Whether a concrete connection between maximal disorder correlations and multifractality exists, and if so under what conditions, remains an open question.

\begin{acknowledgments}
We thank Ivan Khaymovich for helpful comments on the manuscript. This work was supported in part by EPSRC, under Grant No. EP/L015722/1 for the TMCS Centre for Doctoral Training, and Grant No. EP/S020527/1.
\end{acknowledgments}

\bibliography{refs}

\begin{thebibliography}{49}%
\makeatletter
\providecommand \@ifxundefined [1]{%
 \@ifx{#1\undefined}
}%
\providecommand \@ifnum [1]{%
 \ifnum #1\expandafter \@firstoftwo
 \else \expandafter \@secondoftwo
 \fi
}%
\providecommand \@ifx [1]{%
 \ifx #1\expandafter \@firstoftwo
 \else \expandafter \@secondoftwo
 \fi
}%
\providecommand \natexlab [1]{#1}%
\providecommand \enquote  [1]{``#1''}%
\providecommand \bibnamefont  [1]{#1}%
\providecommand \bibfnamefont [1]{#1}%
\providecommand \citenamefont [1]{#1}%
\providecommand \href@noop [0]{\@secondoftwo}%
\providecommand \href [0]{\begingroup \@sanitize@url \@href}%
\providecommand \@href[1]{\@@startlink{#1}\@@href}%
\providecommand \@@href[1]{\endgroup#1\@@endlink}%
\providecommand \@sanitize@url [0]{\catcode `\\12\catcode `\$12\catcode
  `\&12\catcode `\#12\catcode `\^12\catcode `\_12\catcode `\%12\relax}%
\providecommand \@@startlink[1]{}%
\providecommand \@@endlink[0]{}%
\providecommand \url  [0]{\begingroup\@sanitize@url \@url }%
\providecommand \@url [1]{\endgroup\@href {#1}{\urlprefix }}%
\providecommand \urlprefix  [0]{URL }%
\providecommand \Eprint [0]{\href }%
\providecommand \doibase [0]{https://doi.org/}%
\providecommand \selectlanguage [0]{\@gobble}%
\providecommand \bibinfo  [0]{\@secondoftwo}%
\providecommand \bibfield  [0]{\@secondoftwo}%
\providecommand \translation [1]{[#1]}%
\providecommand \BibitemOpen [0]{}%
\providecommand \bibitemStop [0]{}%
\providecommand \bibitemNoStop [0]{.\EOS\space}%
\providecommand \EOS [0]{\spacefactor3000\relax}%
\providecommand \BibitemShut  [1]{\csname bibitem#1\endcsname}%
\let\auto@bib@innerbib\@empty
\bibitem [{\citenamefont {Evers}\ and\ \citenamefont
  {Mirlin}(2008)}]{evers2008anderson}%
  \BibitemOpen
  \bibfield  {author} {\bibinfo {author} {\bibfnamefont {F.}~\bibnamefont
  {Evers}}\ and\ \bibinfo {author} {\bibfnamefont {A.~D.}\ \bibnamefont
  {Mirlin}},\ }\bibfield  {title} {\bibinfo {title} {Anderson transitions},\
  }\href {https://doi.org/10.1103/RevModPhys.80.1355} {\bibfield  {journal}
  {\bibinfo  {journal} {Rev. Mod. Phys.}\ }\textbf {\bibinfo {volume} {80}},\
  \bibinfo {pages} {1355} (\bibinfo {year} {2008})}\BibitemShut {NoStop}%
\bibitem [{\citenamefont {Wegner}(1980)}]{wegner1980inverse}%
  \BibitemOpen
  \bibfield  {author} {\bibinfo {author} {\bibfnamefont {F.}~\bibnamefont
  {Wegner}},\ }\bibfield  {title} {\bibinfo {title} {Inverse participation
  ratio in 2+$\epsilon$ dimensions},\ }\href
  {https://doi.org/10.1007/BF01325284} {\bibfield  {journal} {\bibinfo
  {journal} {Z. Phys. B}\ }\textbf {\bibinfo {volume} {36}},\ \bibinfo {pages}
  {209} (\bibinfo {year} {1980})}\BibitemShut {NoStop}%
\bibitem [{\citenamefont {Castellani}\ and\ \citenamefont
  {Peliti}(1986)}]{castellani1986multifractal}%
  \BibitemOpen
  \bibfield  {author} {\bibinfo {author} {\bibfnamefont {C.}~\bibnamefont
  {Castellani}}\ and\ \bibinfo {author} {\bibfnamefont {L.}~\bibnamefont
  {Peliti}},\ }\href {https://doi.org/10.1088/0305-4470/19/8/004} {\bibfield
  {journal} {\bibinfo  {journal} {Journal of Physics {A}: Mathematical and
  General}\ }\textbf {\bibinfo {volume} {19}},\ \bibinfo {pages} {L429}
  (\bibinfo {year} {1986})}\BibitemShut {NoStop}%
\bibitem [{\citenamefont {Chalker}(1988)}]{chalker1988scaling}%
  \BibitemOpen
  \bibfield  {author} {\bibinfo {author} {\bibfnamefont {J.~T.}\ \bibnamefont
  {Chalker}},\ }\bibfield  {title} {\bibinfo {title} {Scaling and correlations
  at a mobility edge in two dimensions},\ }\href
  {https://iopscience.iop.org/article/10.1088/0022-3719/21/6/004} {\bibfield
  {journal} {\bibinfo  {journal} {J. Phys. C: Solid State Phys.}\ }\textbf
  {\bibinfo {volume} {21}},\ \bibinfo {pages} {L119} (\bibinfo {year}
  {1988})}\BibitemShut {NoStop}%
\bibitem [{\citenamefont {Chalker}(1990)}]{chalker1990scaling}%
  \BibitemOpen
  \bibfield  {author} {\bibinfo {author} {\bibfnamefont {J.~T.}\ \bibnamefont
  {Chalker}},\ }\bibfield  {title} {\bibinfo {title} {Scaling and eigenfunction
  correlations near a mobility edge},\ }\href
  {https://www.sciencedirect.com/science/article/abs/pii/037843719090056X}
  {\bibfield  {journal} {\bibinfo  {journal} {Physica A}\ }\textbf {\bibinfo
  {volume} {167}},\ \bibinfo {pages} {253} (\bibinfo {year}
  {1990})}\BibitemShut {NoStop}%
\bibitem [{\citenamefont {Bauer}\ \emph {et~al.}(1990)\citenamefont {Bauer},
  \citenamefont {Chang},\ and\ \citenamefont {Skinner}}]{bauer1990correlation}%
  \BibitemOpen
  \bibfield  {author} {\bibinfo {author} {\bibfnamefont {J.}~\bibnamefont
  {Bauer}}, \bibinfo {author} {\bibfnamefont {T.-M.}\ \bibnamefont {Chang}},\
  and\ \bibinfo {author} {\bibfnamefont {J.~L.}\ \bibnamefont {Skinner}},\
  }\bibfield  {title} {\bibinfo {title} {Correlation length and
  inverse-participation-ratio exponents and multifractal structure for
  {A}nderson localization},\ }\href {https://doi.org/10.1103/PhysRevB.42.8121}
  {\bibfield  {journal} {\bibinfo  {journal} {Phys. Rev. B}\ }\textbf {\bibinfo
  {volume} {42}},\ \bibinfo {pages} {8121} (\bibinfo {year}
  {1990})}\BibitemShut {NoStop}%
\bibitem [{\citenamefont {Schreiber}\ and\ \citenamefont
  {Grussbach}(1991)}]{schreiber1991multifractal}%
  \BibitemOpen
  \bibfield  {author} {\bibinfo {author} {\bibfnamefont {M.}~\bibnamefont
  {Schreiber}}\ and\ \bibinfo {author} {\bibfnamefont {H.}~\bibnamefont
  {Grussbach}},\ }\bibfield  {title} {\bibinfo {title} {Multifractal wave
  functions at the {A}nderson transition},\ }\href
  {https://doi.org/10.1103/PhysRevLett.67.607} {\bibfield  {journal} {\bibinfo
  {journal} {Phys. Rev. Lett.}\ }\textbf {\bibinfo {volume} {67}},\ \bibinfo
  {pages} {607} (\bibinfo {year} {1991})}\BibitemShut {NoStop}%
\bibitem [{\citenamefont {Janssen}(1994)}]{janssen1994multifractal}%
  \BibitemOpen
  \bibfield  {author} {\bibinfo {author} {\bibfnamefont {M.}~\bibnamefont
  {Janssen}},\ }\bibfield  {title} {\bibinfo {title} {Multifractal analysis of
  broadly-distributed observables at criticality},\ }\href
  {https://doi.org/10.1142/S021797929400049X} {\bibfield  {journal} {\bibinfo
  {journal} {Int. J. Mod. Phys. B}\ }\textbf {\bibinfo {volume} {08}},\
  \bibinfo {pages} {943} (\bibinfo {year} {1994})}\BibitemShut {NoStop}%
\bibitem [{\citenamefont {Chalker}\ \emph {et~al.}(1996)\citenamefont
  {Chalker}, \citenamefont {Kravtsov},\ and\ \citenamefont
  {Lerner}}]{chalker1996spectral}%
  \BibitemOpen
  \bibfield  {author} {\bibinfo {author} {\bibfnamefont {J.~T.}\ \bibnamefont
  {Chalker}}, \bibinfo {author} {\bibfnamefont {V.~E.}\ \bibnamefont
  {Kravtsov}},\ and\ \bibinfo {author} {\bibfnamefont {I.~V.}\ \bibnamefont
  {Lerner}},\ }\bibfield  {title} {\bibinfo {title} {Spectral rigidity and
  eigenfunction correlations at the {A}nderson transition},\ }\href
  {https://doi.org/10.1134/1.567208} {\bibfield  {journal} {\bibinfo  {journal}
  {JETP Letters}\ }\textbf {\bibinfo {volume} {64}},\ \bibinfo {pages} {386}
  (\bibinfo {year} {1996})}\BibitemShut {NoStop}%
\bibitem [{\citenamefont {Mirlin}\ and\ \citenamefont
  {Evers}(2000)}]{mirlin2000multifractality}%
  \BibitemOpen
  \bibfield  {author} {\bibinfo {author} {\bibfnamefont {A.~D.}\ \bibnamefont
  {Mirlin}}\ and\ \bibinfo {author} {\bibfnamefont {F.}~\bibnamefont {Evers}},\
  }\bibfield  {title} {\bibinfo {title} {Multifractality and critical
  fluctuations at the {A}nderson transition},\ }\href
  {https://doi.org/10.1103/PhysRevB.62.7920} {\bibfield  {journal} {\bibinfo
  {journal} {Phys. Rev. B}\ }\textbf {\bibinfo {volume} {62}},\ \bibinfo
  {pages} {7920} (\bibinfo {year} {2000})}\BibitemShut {NoStop}%
\bibitem [{\citenamefont {Evers}\ and\ \citenamefont
  {Mirlin}(2000)}]{evers2000fluc}%
  \BibitemOpen
  \bibfield  {author} {\bibinfo {author} {\bibfnamefont {F.}~\bibnamefont
  {Evers}}\ and\ \bibinfo {author} {\bibfnamefont {A.~D.}\ \bibnamefont
  {Mirlin}},\ }\bibfield  {title} {\bibinfo {title} {Fluctuations of the
  inverse participation ratio at the {A}nderson transition},\ }\href
  {https://doi.org/10.1103/PhysRevLett.84.3690} {\bibfield  {journal} {\bibinfo
   {journal} {Phys. Rev. Lett.}\ }\textbf {\bibinfo {volume} {84}},\ \bibinfo
  {pages} {3690} (\bibinfo {year} {2000})}\BibitemShut {NoStop}%
\bibitem [{\citenamefont {Cuevas}\ \emph
  {et~al.}(2001{\natexlab{a}})\citenamefont {Cuevas}, \citenamefont {Ortu\~no},
  \citenamefont {Gasparian},\ and\ \citenamefont
  {P\'erez-Garrido}}]{cuevas2001fluctuations}%
  \BibitemOpen
  \bibfield  {author} {\bibinfo {author} {\bibfnamefont {E.}~\bibnamefont
  {Cuevas}}, \bibinfo {author} {\bibfnamefont {M.}~\bibnamefont {Ortu\~no}},
  \bibinfo {author} {\bibfnamefont {V.}~\bibnamefont {Gasparian}},\ and\
  \bibinfo {author} {\bibfnamefont {A.}~\bibnamefont {P\'erez-Garrido}},\
  }\bibfield  {title} {\bibinfo {title} {Fluctuations of the correlation
  dimension at metal-insulator transitions},\ }\href
  {https://doi.org/10.1103/PhysRevLett.88.016401} {\bibfield  {journal}
  {\bibinfo  {journal} {Phys. Rev. Lett.}\ }\textbf {\bibinfo {volume} {88}},\
  \bibinfo {pages} {016401} (\bibinfo {year} {2001}{\natexlab{a}})}\BibitemShut
  {NoStop}%
\bibitem [{\citenamefont {Chalker}\ and\ \citenamefont
  {Coddington}(1988)}]{chalker1988percolation}%
  \BibitemOpen
  \bibfield  {author} {\bibinfo {author} {\bibfnamefont {J.~T.}\ \bibnamefont
  {Chalker}}\ and\ \bibinfo {author} {\bibfnamefont {P.~D.}\ \bibnamefont
  {Coddington}},\ }\bibfield  {title} {\bibinfo {title} {Percolation, quantum
  tunnelling and the integer {H}all effect},\ }\href
  {http://stacks.iop.org/0022-3719/21/i=14/a=008} {\bibfield  {journal}
  {\bibinfo  {journal} {J. Phys. C: Solid State Phys.}\ }\textbf {\bibinfo
  {volume} {21}},\ \bibinfo {pages} {2665} (\bibinfo {year}
  {1988})}\BibitemShut {NoStop}%
\bibitem [{\citenamefont {Huckestein}(1995)}]{huckenstein1995scaling}%
  \BibitemOpen
  \bibfield  {author} {\bibinfo {author} {\bibfnamefont {B.}~\bibnamefont
  {Huckestein}},\ }\bibfield  {title} {\bibinfo {title} {Scaling theory of the
  integer quantum {H}all effect},\ }\href
  {https://doi.org/10.1103/RevModPhys.67.357} {\bibfield  {journal} {\bibinfo
  {journal} {Rev. Mod. Phys.}\ }\textbf {\bibinfo {volume} {67}},\ \bibinfo
  {pages} {357} (\bibinfo {year} {1995})}\BibitemShut {NoStop}%
\bibitem [{\citenamefont {Evers}\ \emph {et~al.}(2001)\citenamefont {Evers},
  \citenamefont {Mildenberger},\ and\ \citenamefont
  {Mirlin}}]{evers2001multifractality}%
  \BibitemOpen
  \bibfield  {author} {\bibinfo {author} {\bibfnamefont {F.}~\bibnamefont
  {Evers}}, \bibinfo {author} {\bibfnamefont {A.}~\bibnamefont
  {Mildenberger}},\ and\ \bibinfo {author} {\bibfnamefont {A.~D.}\ \bibnamefont
  {Mirlin}},\ }\bibfield  {title} {\bibinfo {title} {Multifractality of wave
  functions at the quantum {H}all transition revisited},\ }\href
  {https://doi.org/10.1103/PhysRevB.64.241303} {\bibfield  {journal} {\bibinfo
  {journal} {Phys. Rev. B}\ }\textbf {\bibinfo {volume} {64}},\ \bibinfo
  {pages} {241303} (\bibinfo {year} {2001})}\BibitemShut {NoStop}%
\bibitem [{\citenamefont {Levitov}(1990)}]{levitov1990delocalisation}%
  \BibitemOpen
  \bibfield  {author} {\bibinfo {author} {\bibfnamefont {L.~S.}\ \bibnamefont
  {Levitov}},\ }\bibfield  {title} {\bibinfo {title} {Delocalization of
  vibrational modes caused by electric dipole interaction},\ }\href
  {https://doi.org/10.1103/PhysRevLett.64.547} {\bibfield  {journal} {\bibinfo
  {journal} {Phys. Rev. Lett.}\ }\textbf {\bibinfo {volume} {64}},\ \bibinfo
  {pages} {547} (\bibinfo {year} {1990})}\BibitemShut {NoStop}%
\bibitem [{\citenamefont {Mirlin}\ \emph {et~al.}(1996)\citenamefont {Mirlin},
  \citenamefont {Fyodorov}, \citenamefont {Dittes}, \citenamefont {Quezada},\
  and\ \citenamefont {Seligman}}]{mirlin1996transition}%
  \BibitemOpen
  \bibfield  {author} {\bibinfo {author} {\bibfnamefont {A.~D.}\ \bibnamefont
  {Mirlin}}, \bibinfo {author} {\bibfnamefont {Y.~V.}\ \bibnamefont
  {Fyodorov}}, \bibinfo {author} {\bibfnamefont {F.-M.}\ \bibnamefont
  {Dittes}}, \bibinfo {author} {\bibfnamefont {J.}~\bibnamefont {Quezada}},\
  and\ \bibinfo {author} {\bibfnamefont {T.~H.}\ \bibnamefont {Seligman}},\
  }\bibfield  {title} {\bibinfo {title} {Transition from localized to extended
  eigenstates in the ensemble of power-law random banded matrices},\ }\href
  {https://doi.org/10.1103/PhysRevE.54.3221} {\bibfield  {journal} {\bibinfo
  {journal} {Phys. Rev. E}\ }\textbf {\bibinfo {volume} {54}},\ \bibinfo
  {pages} {3221} (\bibinfo {year} {1996})}\BibitemShut {NoStop}%
\bibitem [{\citenamefont {Parshin}\ and\ \citenamefont
  {Schober}(1998)}]{parshin1998multifractal}%
  \BibitemOpen
  \bibfield  {author} {\bibinfo {author} {\bibfnamefont {D.~A.}\ \bibnamefont
  {Parshin}}\ and\ \bibinfo {author} {\bibfnamefont {H.~R.}\ \bibnamefont
  {Schober}},\ }\bibfield  {title} {\bibinfo {title} {Multifractal structure of
  eigenstates in the {A}nderson model with long-range off-diagonal disorder},\
  }\href {https://doi.org/10.1103/PhysRevB.57.10232} {\bibfield  {journal}
  {\bibinfo  {journal} {Phys. Rev. B}\ }\textbf {\bibinfo {volume} {57}},\
  \bibinfo {pages} {10232} (\bibinfo {year} {1998})}\BibitemShut {NoStop}%
\bibitem [{\citenamefont {Levitov}(1999)}]{levitov1999critical}%
  \BibitemOpen
  \bibfield  {author} {\bibinfo {author} {\bibfnamefont {L.~S.}\ \bibnamefont
  {Levitov}},\ }\bibfield  {title} {\bibinfo {title} {Critical {H}amiltonians
  with long range hopping},\ }\href
  {https://onlinelibrary.wiley.com/doi/10.1002/(SICI)1521-3889(199911)8:7/9%3C697::AID-ANDP697%3E3.0.CO;2-W}
  {\bibfield  {journal} {\bibinfo  {journal} {Annalen der Physik}\ }\textbf
  {\bibinfo {volume} {8}},\ \bibinfo {pages} {697} (\bibinfo {year}
  {1999})}\BibitemShut {NoStop}%
\bibitem [{\citenamefont {Varga}\ and\ \citenamefont
  {Braun}(2000)}]{varga2000critical}%
  \BibitemOpen
  \bibfield  {author} {\bibinfo {author} {\bibfnamefont {I.}~\bibnamefont
  {Varga}}\ and\ \bibinfo {author} {\bibfnamefont {D.}~\bibnamefont {Braun}},\
  }\bibfield  {title} {\bibinfo {title} {Critical statistics in a power-law
  random-banded matrix ensemble},\ }\href
  {https://doi.org/10.1103/PhysRevB.61.R11859} {\bibfield  {journal} {\bibinfo
  {journal} {Phys. Rev. B}\ }\textbf {\bibinfo {volume} {61}},\ \bibinfo
  {pages} {R11859} (\bibinfo {year} {2000})}\BibitemShut {NoStop}%
\bibitem [{\citenamefont {Cuevas}\ \emph
  {et~al.}(2001{\natexlab{b}})\citenamefont {Cuevas}, \citenamefont
  {Gasparian},\ and\ \citenamefont {Ortu\~no}}]{cuevas2001anomalously}%
  \BibitemOpen
  \bibfield  {author} {\bibinfo {author} {\bibfnamefont {E.}~\bibnamefont
  {Cuevas}}, \bibinfo {author} {\bibfnamefont {V.}~\bibnamefont {Gasparian}},\
  and\ \bibinfo {author} {\bibfnamefont {M.}~\bibnamefont {Ortu\~no}},\
  }\bibfield  {title} {\bibinfo {title} {Anomalously large critical regions in
  power-law random matrix ensembles},\ }\href
  {https://doi.org/10.1103/PhysRevLett.87.056601} {\bibfield  {journal}
  {\bibinfo  {journal} {Phys. Rev. Lett.}\ }\textbf {\bibinfo {volume} {87}},\
  \bibinfo {pages} {056601} (\bibinfo {year} {2001}{\natexlab{b}})}\BibitemShut
  {NoStop}%
\bibitem [{\citenamefont {Cuevas}(2003)}]{cuevas2003multifractality}%
  \BibitemOpen
  \bibfield  {author} {\bibinfo {author} {\bibfnamefont {E.}~\bibnamefont
  {Cuevas}},\ }\bibfield  {title} {\bibinfo {title} {Multifractality of
  {H}amiltonians with power-law transfer terms},\ }\href
  {https://doi.org/10.1103/PhysRevB.68.184206} {\bibfield  {journal} {\bibinfo
  {journal} {Phys. Rev. B}\ }\textbf {\bibinfo {volume} {68}},\ \bibinfo
  {pages} {184206} (\bibinfo {year} {2003})}\BibitemShut {NoStop}%
\bibitem [{\citenamefont {Monthus}\ and\ \citenamefont
  {Garel}(2010)}]{monthus2010anderson}%
  \BibitemOpen
  \bibfield  {author} {\bibinfo {author} {\bibfnamefont {C.}~\bibnamefont
  {Monthus}}\ and\ \bibinfo {author} {\bibfnamefont {T.}~\bibnamefont
  {Garel}},\ }\bibfield  {title} {\bibinfo {title} {The {A}nderson localization
  transition with long-ranged hoppings: analysis of the strong multifractality
  regime in terms of weighted {L}{\'e}vy sums},\ }\href
  {https://iopscience.iop.org/article/10.1088/1742-5468/2010/09/P09015}
  {\bibfield  {journal} {\bibinfo  {journal} {J. Stat. Mech.}\ ,\ \bibinfo
  {pages} {P09015}} (\bibinfo {year} {2010})}\BibitemShut {NoStop}%
\bibitem [{\citenamefont {Kravtsov}\ \emph {et~al.}(2015)\citenamefont
  {Kravtsov}, \citenamefont {Khaymovich}, \citenamefont {Cuevas},\ and\
  \citenamefont {Amini}}]{kravtsov2015random}%
  \BibitemOpen
  \bibfield  {author} {\bibinfo {author} {\bibfnamefont {V.~E.}\ \bibnamefont
  {Kravtsov}}, \bibinfo {author} {\bibfnamefont {I.~M.}\ \bibnamefont
  {Khaymovich}}, \bibinfo {author} {\bibfnamefont {E.}~\bibnamefont {Cuevas}},\
  and\ \bibinfo {author} {\bibfnamefont {M.}~\bibnamefont {Amini}},\ }\bibfield
   {title} {\bibinfo {title} {A random matrix model with localization and
  ergodic transitions},\ }\href
  {https://doi.org/10.1088/1367-2630/17/12/122002} {\bibfield  {journal}
  {\bibinfo  {journal} {New Journal of Physics}\ }\textbf {\bibinfo {volume}
  {17}},\ \bibinfo {pages} {122002} (\bibinfo {year} {2015})}\BibitemShut
  {NoStop}%
\bibitem [{\citenamefont {Deng}\ \emph {et~al.}(2019)\citenamefont {Deng},
  \citenamefont {Ray}, \citenamefont {Sinha}, \citenamefont {Shlyapnikov},\
  and\ \citenamefont {Santos}}]{deng2019one}%
  \BibitemOpen
  \bibfield  {author} {\bibinfo {author} {\bibfnamefont {X.}~\bibnamefont
  {Deng}}, \bibinfo {author} {\bibfnamefont {S.}~\bibnamefont {Ray}}, \bibinfo
  {author} {\bibfnamefont {S.}~\bibnamefont {Sinha}}, \bibinfo {author}
  {\bibfnamefont {G.~V.}\ \bibnamefont {Shlyapnikov}},\ and\ \bibinfo {author}
  {\bibfnamefont {L.}~\bibnamefont {Santos}},\ }\bibfield  {title} {\bibinfo
  {title} {One-dimensional quasicrystals with power-law hopping},\ }\href
  {https://doi.org/10.1103/PhysRevLett.123.025301} {\bibfield  {journal}
  {\bibinfo  {journal} {Phys. Rev. Lett.}\ }\textbf {\bibinfo {volume} {123}},\
  \bibinfo {pages} {025301} (\bibinfo {year} {2019})}\BibitemShut {NoStop}%
\bibitem [{\citenamefont {Monthus}(2019)}]{monthus2019multifractality}%
  \BibitemOpen
  \bibfield  {author} {\bibinfo {author} {\bibfnamefont {C.}~\bibnamefont
  {Monthus}},\ }\bibfield  {title} {\bibinfo {title} {Multifractality in the
  generalized {A}ubry–{A}ndr{\'e} quasiperiodic localization model with
  power-law hoppings or power-law {F}ourier coefficients},\ }\href
  {https://doi.org/10.1142/S0218348X19500075} {\bibfield  {journal} {\bibinfo
  {journal} {Fractals}\ }\textbf {\bibinfo {volume} {27}},\ \bibinfo {pages}
  {1950007} (\bibinfo {year} {2019})}\BibitemShut {NoStop}%
\bibitem [{\citenamefont {Khaymovich}\ \emph {et~al.}(2020)\citenamefont
  {Khaymovich}, \citenamefont {Kravtsov}, \citenamefont {Altshuler},\ and\
  \citenamefont {Ioffe}}]{khaymovich2020fragile}%
  \BibitemOpen
  \bibfield  {author} {\bibinfo {author} {\bibfnamefont {I.~M.}\ \bibnamefont
  {Khaymovich}}, \bibinfo {author} {\bibfnamefont {V.~E.}\ \bibnamefont
  {Kravtsov}}, \bibinfo {author} {\bibfnamefont {B.~L.}\ \bibnamefont
  {Altshuler}},\ and\ \bibinfo {author} {\bibfnamefont {L.~B.}\ \bibnamefont
  {Ioffe}},\ }\bibfield  {title} {\bibinfo {title} {Fragile extended phases in
  the log-normal {R}osenzweig-{P}orter model},\ }\href
  {https://doi.org/10.1103/PhysRevResearch.2.043346} {\bibfield  {journal}
  {\bibinfo  {journal} {Phys. Rev. Research}\ }\textbf {\bibinfo {volume}
  {2}},\ \bibinfo {pages} {043346} (\bibinfo {year} {2020})}\BibitemShut
  {NoStop}%
\bibitem [{\citenamefont {Kravtsov}\ \emph {et~al.}(2020)\citenamefont
  {Kravtsov}, \citenamefont {Khaymovich}, \citenamefont {Altshuler},\ and\
  \citenamefont {Ioffe}}]{kravtsov2020localization}%
  \BibitemOpen
  \bibfield  {author} {\bibinfo {author} {\bibfnamefont {V.~E.}\ \bibnamefont
  {Kravtsov}}, \bibinfo {author} {\bibfnamefont {I.~M.}\ \bibnamefont
  {Khaymovich}}, \bibinfo {author} {\bibfnamefont {B.~L.}\ \bibnamefont
  {Altshuler}},\ and\ \bibinfo {author} {\bibfnamefont {L.~B.}\ \bibnamefont
  {Ioffe}},\ }\href@noop {} {\bibinfo {title} {Localization transition on the
  random regular graph as an unstable tricritical point in a log-normal
  rosenzweig-porter random matrix ensemble}} (\bibinfo {year} {2020}),\ \Eprint
  {https://arxiv.org/abs/2002.02979} {arXiv:2002.02979 [cond-mat.dis-nn]}
  \BibitemShut {NoStop}%
\bibitem [{\citenamefont {Biroli}\ and\ \citenamefont
  {Tarzia}(2021)}]{biroli2021levy}%
  \BibitemOpen
  \bibfield  {author} {\bibinfo {author} {\bibfnamefont {G.}~\bibnamefont
  {Biroli}}\ and\ \bibinfo {author} {\bibfnamefont {M.}~\bibnamefont
  {Tarzia}},\ }\bibfield  {title} {\bibinfo {title}
  {L{\'e}vy-{R}osenzweig-{P}orter random matrix ensemble},\ }\href
  {https://doi.org/10.1103/PhysRevB.103.104205} {\bibfield  {journal} {\bibinfo
   {journal} {Phys. Rev. B}\ }\textbf {\bibinfo {volume} {103}},\ \bibinfo
  {pages} {104205} (\bibinfo {year} {2021})}\BibitemShut {NoStop}%
\bibitem [{\citenamefont {Roy}\ \emph {et~al.}(2018)\citenamefont {Roy},
  \citenamefont {Khaymovich}, \citenamefont {Das},\ and\ \citenamefont
  {Moessner}}]{roy2018multifractality}%
  \BibitemOpen
  \bibfield  {author} {\bibinfo {author} {\bibfnamefont {S.}~\bibnamefont
  {Roy}}, \bibinfo {author} {\bibfnamefont {I.~M.}\ \bibnamefont {Khaymovich}},
  \bibinfo {author} {\bibfnamefont {A.}~\bibnamefont {Das}},\ and\ \bibinfo
  {author} {\bibfnamefont {R.}~\bibnamefont {Moessner}},\ }\bibfield  {title}
  {\bibinfo {title} {{Multifractality without fine-tuning in a Floquet
  quasiperiodic chain}},\ }\href {https://doi.org/10.21468/SciPostPhys.4.5.025}
  {\bibfield  {journal} {\bibinfo  {journal} {SciPost Phys.}\ }\textbf
  {\bibinfo {volume} {4}},\ \bibinfo {pages} {25} (\bibinfo {year}
  {2018})}\BibitemShut {NoStop}%
\bibitem [{\citenamefont {Sarkar}\ \emph {et~al.}(2021)\citenamefont {Sarkar},
  \citenamefont {Ghosh}, \citenamefont {Sen},\ and\ \citenamefont
  {Sengupta}}]{sarkar2021mobility}%
  \BibitemOpen
  \bibfield  {author} {\bibinfo {author} {\bibfnamefont {M.}~\bibnamefont
  {Sarkar}}, \bibinfo {author} {\bibfnamefont {R.}~\bibnamefont {Ghosh}},
  \bibinfo {author} {\bibfnamefont {A.}~\bibnamefont {Sen}},\ and\ \bibinfo
  {author} {\bibfnamefont {K.}~\bibnamefont {Sengupta}},\ }\bibfield  {title}
  {\bibinfo {title} {Mobility edge and multifractality in a periodically driven
  {A}ubry-{A}ndr\'e model},\ }\href
  {https://doi.org/10.1103/PhysRevB.103.184309} {\bibfield  {journal} {\bibinfo
   {journal} {Phys. Rev. B}\ }\textbf {\bibinfo {volume} {103}},\ \bibinfo
  {pages} {184309} (\bibinfo {year} {2021})}\BibitemShut {NoStop}%
\bibitem [{\citenamefont {De~Tomasi}\ \emph {et~al.}(2016)\citenamefont
  {De~Tomasi}, \citenamefont {Roy},\ and\ \citenamefont
  {Bera}}]{detomasi2016generalised}%
  \BibitemOpen
  \bibfield  {author} {\bibinfo {author} {\bibfnamefont {G.}~\bibnamefont
  {De~Tomasi}}, \bibinfo {author} {\bibfnamefont {S.}~\bibnamefont {Roy}},\
  and\ \bibinfo {author} {\bibfnamefont {S.}~\bibnamefont {Bera}},\ }\bibfield
  {title} {\bibinfo {title} {Generalized {D}yson model: {N}ature of the zero
  mode and its implication in dynamics},\ }\href
  {https://doi.org/10.1103/PhysRevB.94.144202} {\bibfield  {journal} {\bibinfo
  {journal} {Phys. Rev. B}\ }\textbf {\bibinfo {volume} {94}},\ \bibinfo
  {pages} {144202} (\bibinfo {year} {2016})}\BibitemShut {NoStop}%
\bibitem [{\citenamefont {Monthus}(2017)}]{monthus2017multifractality}%
  \BibitemOpen
  \bibfield  {author} {\bibinfo {author} {\bibfnamefont {C.}~\bibnamefont
  {Monthus}},\ }\bibfield  {title} {\bibinfo {title} {Multifractality of
  eigenstates in the delocalized non-ergodic phase of some random matrix
  models: {W}igner{\textendash}{W}eisskopf approach},\ }\href
  {https://doi.org/10.1088/1751-8121/aa77e1} {\bibfield  {journal} {\bibinfo
  {journal} {J. Phys. A: Math. Theor.}\ }\textbf {\bibinfo {volume} {50}},\
  \bibinfo {pages} {295101} (\bibinfo {year} {2017})}\BibitemShut {NoStop}%
\bibitem [{\citenamefont {Khaymovich}\ and\ \citenamefont
  {Kravtsov}(2021)}]{khyamovich2021dynamical}%
  \BibitemOpen
  \bibfield  {author} {\bibinfo {author} {\bibfnamefont {I.~M.}\ \bibnamefont
  {Khaymovich}}\ and\ \bibinfo {author} {\bibfnamefont {V.~E.}\ \bibnamefont
  {Kravtsov}},\ }\bibfield  {title} {\bibinfo {title} {{Dynamical phases in a
  ``multifractal'' {R}osenzweig-{P}orter model}},\ }\href
  {https://doi.org/10.21468/SciPostPhys.11.2.045} {\bibfield  {journal}
  {\bibinfo  {journal} {SciPost Phys.}\ }\textbf {\bibinfo {volume} {11}},\
  \bibinfo {pages} {45} (\bibinfo {year} {2021})}\BibitemShut {NoStop}%
\bibitem [{Note1()}]{Note1}%
  \BibitemOpen
  \bibinfo {note} {It should be noted that multifractality is not always a
  prerequisite for slow dynamics~\cite
  {detomasi2020subdiffusion,khyamovich2021dynamical}.}\BibitemShut {Stop}%
\bibitem [{\citenamefont {Roy}\ and\ \citenamefont
  {Logan}(2020{\natexlab{a}})}]{roy2020correlations}%
  \BibitemOpen
  \bibfield  {author} {\bibinfo {author} {\bibfnamefont {S.}~\bibnamefont
  {Roy}}\ and\ \bibinfo {author} {\bibfnamefont {D.~E.}\ \bibnamefont
  {Logan}},\ }\bibfield  {title} {\bibinfo {title} {Fock-space correlations and
  the origins of many-body localization},\ }\href
  {https://doi.org/10.1103/PhysRevB.101.134202} {\bibfield  {journal} {\bibinfo
   {journal} {Phys. Rev. B}\ }\textbf {\bibinfo {volume} {101}},\ \bibinfo
  {pages} {134202} (\bibinfo {year} {2020}{\natexlab{a}})}\BibitemShut
  {NoStop}%
\bibitem [{\citenamefont {Roy}\ and\ \citenamefont
  {Logan}(2020{\natexlab{b}})}]{roy2020strongly}%
  \BibitemOpen
  \bibfield  {author} {\bibinfo {author} {\bibfnamefont {S.}~\bibnamefont
  {Roy}}\ and\ \bibinfo {author} {\bibfnamefont {D.~E.}\ \bibnamefont
  {Logan}},\ }\bibfield  {title} {\bibinfo {title} {Localization on certain
  graphs with strongly correlated disorder},\ }\href
  {https://doi.org/10.1103/PhysRevLett.125.250402} {\bibfield  {journal}
  {\bibinfo  {journal} {Phys. Rev. Lett.}\ }\textbf {\bibinfo {volume} {125}},\
  \bibinfo {pages} {250402} (\bibinfo {year} {2020}{\natexlab{b}})}\BibitemShut
  {NoStop}%
\bibitem [{\citenamefont {Anderson}(1958)}]{anderson1958absence}%
  \BibitemOpen
  \bibfield  {author} {\bibinfo {author} {\bibfnamefont {P.~W.}\ \bibnamefont
  {Anderson}},\ }\bibfield  {title} {\bibinfo {title} {Absence of diffusion in
  certain random lattices},\ }\href {https://doi.org/10.1103/PhysRev.109.1492}
  {\bibfield  {journal} {\bibinfo  {journal} {Phys. Rev.}\ }\textbf {\bibinfo
  {volume} {109}},\ \bibinfo {pages} {1492} (\bibinfo {year}
  {1958})}\BibitemShut {NoStop}%
\bibitem [{\citenamefont {Mott}\ and\ \citenamefont
  {Twose}(1961)}]{mott1961theory}%
  \BibitemOpen
  \bibfield  {author} {\bibinfo {author} {\bibfnamefont {N.~F.}\ \bibnamefont
  {Mott}}\ and\ \bibinfo {author} {\bibfnamefont {W.~D.}\ \bibnamefont
  {Twose}},\ }\bibfield  {title} {\bibinfo {title} {The theory of impurity
  conduction},\ }\href {https://doi.org/10.1080/00018736100101271} {\bibfield
  {journal} {\bibinfo  {journal} {Advances in Physics}\ }\textbf {\bibinfo
  {volume} {10}},\ \bibinfo {pages} {107} (\bibinfo {year} {1961})}\BibitemShut
  {NoStop}%
\bibitem [{sup()}]{supp}%
  \BibitemOpen
  \href@noop {} {}\bibinfo {note} {See supplementary material at [URL] for (I)
  results on transmittance and (II) derivation of the lack of self-averaging in
  the density of states.}\BibitemShut {Stop}%
\bibitem [{Note2()}]{Note2}%
  \BibitemOpen
  \bibinfo {note} {As each tabletop state has two breaks, and the density of
  tabletops of length $\ell \sim L^{\alpha }$ is $\sim L^{-\alpha
  }$.}\BibitemShut {Stop}%
\bibitem [{Note3()}]{Note3}%
  \BibitemOpen
  \bibinfo {note} {Perhaps expectedly, given that the Hausdorff dimension of
  any tree with $K\geq 2$ is infinite, while that of a 1D chain is
  unity.}\BibitemShut {Stop}%
\bibitem [{\citenamefont {Izrailev}\ and\ \citenamefont
  {Krokhin}(1999)}]{izrailev1999localisation}%
  \BibitemOpen
  \bibfield  {author} {\bibinfo {author} {\bibfnamefont {F.~M.}\ \bibnamefont
  {Izrailev}}\ and\ \bibinfo {author} {\bibfnamefont {A.~A.}\ \bibnamefont
  {Krokhin}},\ }\bibfield  {title} {\bibinfo {title} {Localization and the
  mobility edge in one-dimensional potentials with correlated disorder},\
  }\href {https://doi.org/10.1103/PhysRevLett.82.4062} {\bibfield  {journal}
  {\bibinfo  {journal} {Phys. Rev. Lett.}\ }\textbf {\bibinfo {volume} {82}},\
  \bibinfo {pages} {4062} (\bibinfo {year} {1999})}\BibitemShut {NoStop}%
\bibitem [{\citenamefont {Croy}\ \emph {et~al.}(2011)\citenamefont {Croy},
  \citenamefont {Cain},\ and\ \citenamefont {Schreiber}}]{croy2011anderson}%
  \BibitemOpen
  \bibfield  {author} {\bibinfo {author} {\bibfnamefont {A.}~\bibnamefont
  {Croy}}, \bibinfo {author} {\bibfnamefont {P.}~\bibnamefont {Cain}},\ and\
  \bibinfo {author} {\bibfnamefont {M.}~\bibnamefont {Schreiber}},\ }\bibfield
  {title} {\bibinfo {title} {Anderson localization in 1d systems with
  correlated disorder},\ }\href {https://doi.org/10.1140/epjb/e2011-20212-1}
  {\bibfield  {journal} {\bibinfo  {journal} {Eur. Phys. J. B}\ }\textbf
  {\bibinfo {volume} {82}},\ \bibinfo {pages} {107} (\bibinfo {year}
  {2011})}\BibitemShut {NoStop}%
\bibitem [{\citenamefont {De~Luca}\ and\ \citenamefont
  {Scardicchio}(2013)}]{deluca2013ergodicity}%
  \BibitemOpen
  \bibfield  {author} {\bibinfo {author} {\bibfnamefont {A.}~\bibnamefont
  {De~Luca}}\ and\ \bibinfo {author} {\bibfnamefont {A.}~\bibnamefont
  {Scardicchio}},\ }\bibfield  {title} {\bibinfo {title} {Ergodicity breaking
  in a model showing many-body localization},\ }\href
  {https://doi.org/10.1209/0295-5075/101/37003} {\bibfield  {journal} {\bibinfo
   {journal} {Europhys. Lett.}\ }\textbf {\bibinfo {volume} {101}},\ \bibinfo
  {pages} {37003} (\bibinfo {year} {2013})}\BibitemShut {NoStop}%
\bibitem [{\citenamefont {Mac\'e}\ \emph {et~al.}(2019)\citenamefont {Mac\'e},
  \citenamefont {Alet},\ and\ \citenamefont
  {Laflorencie}}]{mace2018multifractal}%
  \BibitemOpen
  \bibfield  {author} {\bibinfo {author} {\bibfnamefont {N.}~\bibnamefont
  {Mac\'e}}, \bibinfo {author} {\bibfnamefont {F.}~\bibnamefont {Alet}},\ and\
  \bibinfo {author} {\bibfnamefont {N.}~\bibnamefont {Laflorencie}},\
  }\bibfield  {title} {\bibinfo {title} {Multifractal scalings across the
  many-body localization transition},\ }\href
  {https://doi.org/10.1103/PhysRevLett.123.180601} {\bibfield  {journal}
  {\bibinfo  {journal} {Phys. Rev. Lett.}\ }\textbf {\bibinfo {volume} {123}},\
  \bibinfo {pages} {180601} (\bibinfo {year} {2019})}\BibitemShut {NoStop}%
\bibitem [{\citenamefont {De~Tomasi}\ \emph {et~al.}(2021)\citenamefont
  {De~Tomasi}, \citenamefont {Khaymovich}, \citenamefont {Pollmann},\ and\
  \citenamefont {Warzel}}]{detomasi2021PRB}%
  \BibitemOpen
  \bibfield  {author} {\bibinfo {author} {\bibfnamefont {G.}~\bibnamefont
  {De~Tomasi}}, \bibinfo {author} {\bibfnamefont {I.~M.}\ \bibnamefont
  {Khaymovich}}, \bibinfo {author} {\bibfnamefont {F.}~\bibnamefont
  {Pollmann}},\ and\ \bibinfo {author} {\bibfnamefont {S.}~\bibnamefont
  {Warzel}},\ }\bibfield  {title} {\bibinfo {title} {Rare thermal bubbles at
  the many-body localization transition from the {F}ock space point of view},\
  }\href {https://doi.org/10.1103/PhysRevB.104.024202} {\bibfield  {journal}
  {\bibinfo  {journal} {Phys. Rev. B}\ }\textbf {\bibinfo {volume} {104}},\
  \bibinfo {pages} {024202} (\bibinfo {year} {2021})}\BibitemShut {NoStop}%
\bibitem [{\citenamefont {Roy}\ and\ \citenamefont
  {Logan}(2021)}]{roy2021fockspace}%
  \BibitemOpen
  \bibfield  {author} {\bibinfo {author} {\bibfnamefont {S.}~\bibnamefont
  {Roy}}\ and\ \bibinfo {author} {\bibfnamefont {D.~E.}\ \bibnamefont
  {Logan}},\ }\bibfield  {title} {\bibinfo {title} {Fock-space anatomy of
  eigenstates across the many-body localization transition},\ }\href
  {https://doi.org/10.1103/PhysRevB.104.174201} {\bibfield  {journal} {\bibinfo
   {journal} {Phys. Rev. B}\ }\textbf {\bibinfo {volume} {104}},\ \bibinfo
  {pages} {174201} (\bibinfo {year} {2021})}\BibitemShut {NoStop}%
\bibitem [{\citenamefont {De~Tomasi}\ \emph {et~al.}(2020)\citenamefont
  {De~Tomasi}, \citenamefont {Bera}, \citenamefont {Scardicchio},\ and\
  \citenamefont {Khaymovich}}]{detomasi2020subdiffusion}%
  \BibitemOpen
  \bibfield  {author} {\bibinfo {author} {\bibfnamefont {G.}~\bibnamefont
  {De~Tomasi}}, \bibinfo {author} {\bibfnamefont {S.}~\bibnamefont {Bera}},
  \bibinfo {author} {\bibfnamefont {A.}~\bibnamefont {Scardicchio}},\ and\
  \bibinfo {author} {\bibfnamefont {I.~M.}\ \bibnamefont {Khaymovich}},\
  }\bibfield  {title} {\bibinfo {title} {Subdiffusion in the anderson model on
  the random regular graph},\ }\href
  {https://doi.org/10.1103/PhysRevB.101.100201} {\bibfield  {journal} {\bibinfo
   {journal} {Phys. Rev. B}\ }\textbf {\bibinfo {volume} {101}},\ \bibinfo
  {pages} {100201} (\bibinfo {year} {2020})}\BibitemShut {NoStop}%
\end{thebibliography}%

\clearpage


\section*{Supplementary material}

# Supplementary material: Anomalous multifractality in quantum chains with strongly correlated disorder

Alexander Duthie, Sthitadhi Roy, and David E. Logan

This supplementary material contains two sections. Sec. I shows results for the transmittance, computed using a numerical transfer matrix approach. Sec. II presents analytic results for the lack of self-averaging in the density of states.

## I. TRANSFER MATRIX

The transmittance $T$ of an incident wave through a finite-size system probes the localisation properties of the underlying eigenstates. To compute the transmittance, we envisage the correlated chain with sites labelled $x=1$ through $x=L$ being connected to semi-infinite clean chains on either end. The incident wavefunction is a Bloch wave, with $\psi(x) = e^{ikx}$ for $x \leq 1$ and energy $E = 2\cos k$. The transmittance $T$ is then given via the relation $|\psi(L)|^2 = T|\psi(1)|^2$.

Realisations hosting extended eigenstates spanning the entire chain at energy $E$ are expected to yield a finite transmittance in the thermodynamic limit. By contrast, if the eigenstates are multifractal and subextensive, the transmittance is vanishing. This reflects the tabletop nature of the eigenstates, which have vanishing amplitude on sites outside the tabletops. From the point of view of transmittances, subextensive tabletop states are more akin to localised states, and markedly different from conventional multifractal states which, due to their long polynomial tails, have transmittances which decay anomalously with system size.

We compute the transmittance using a numerical transfer matrix approach, which involves propagating the incident wavefunction recursively as

$$\begin{bmatrix} \psi(x+1) \\ \psi(x) \end{bmatrix} = \begin{bmatrix} W\epsilon_x - E & -1 \\ 1 & 0 \end{bmatrix} \begin{bmatrix} \psi(x) \\ \psi(x-1) \end{bmatrix}, \quad (S1)$$

where $E$ is the energy of the incident wavefunction and $\{\epsilon_x\}$ is the set of correlated onsite energies. The results are shown in Fig. S1. In the left panel, we plot the mean and typical transmittances $T^{m/t}$ as a function of $W$, for incident waves of energy 0 (with Hamiltonian $\mathcal{H} = H - \mathrm{Tr}H$). The mean transmittance vanishes when *all* realisations host multifractal states; the typical transmittance, on the other hand, vanishes when even one realisation admits a zero transmittance. The mixed regime thus exists between the vanishing of these two curves.

Fig. S1 also shows distributions $P_T(T)$ of transmittances. In the extended phase ($W=0.5$), $P_T(T)$ is supported strictly on $T>0$, as expected for conducting states. In the mixed regime ($W=1.5$), a broad distribution of finite transmittances is seen, corresponding to conducting states, alongside a peak at $T=0$, corresponding to the tabletop multifractal states. For the phase with only multifractal states ($W=5$), only the peak at $T=0$ arises.

## II. FLUCTUATIONS IN THE DENSITY OF STATES

In this section we derive results showing the lack of self-averaging in the density of states. The mean eigenvalue of the Hamiltonian in Eq. (1) of the main text is $\mathrm{Tr}H = L^{-1}\sum_n E_n$, where $H|\psi_n\rangle = E_n|\psi_n\rangle$. Changing to the basis of sites, we write this as $\mathrm{Tr}H = L^{-1}\sum_x W\epsilon_x$. Defining

$$S_1 = W^{-1}\mathrm{Tr}H = \frac{1}{L}\sum_x \epsilon_x, \quad (S2)$$

we have $\langle S_1 \rangle = 0$. As a sum of Gaussian random variables, $S_1$ is itself Gaussian distributed, with variance given by

$$L^2\langle S_1^2 \rangle = \sum_{x,y} \langle \epsilon_x \epsilon_y \rangle \quad (S3)$$

With $\langle \epsilon_x \epsilon_y \rangle = \exp[-|x-y|/L]$, $\langle S_1^2 \rangle = 2/e + \mathcal{O}(L^{-1})$. The mean eigenvalue thus has a nonzero variance in the thermodynamic limit and hence fluctuates from realisation-to-realisation, which necessitates the recentring of the Hamiltonian $H \to \mathcal{H} = H - \mathrm{Tr}H$.

The variance of the spectrum of a single disorder realisation of the *re-centred* Hamiltonian $H - \mathrm{Tr}H$ is given by

$$\mu_E^2 = \mathrm{Tr}(H - \mathrm{Tr}(H))^2 = \frac{W^2}{L}\sum_x \epsilon_x^2 - \frac{W^2}{L^2}\sum_{x\neq y}\epsilon_x \epsilon_y + 2. \quad (S4)$$

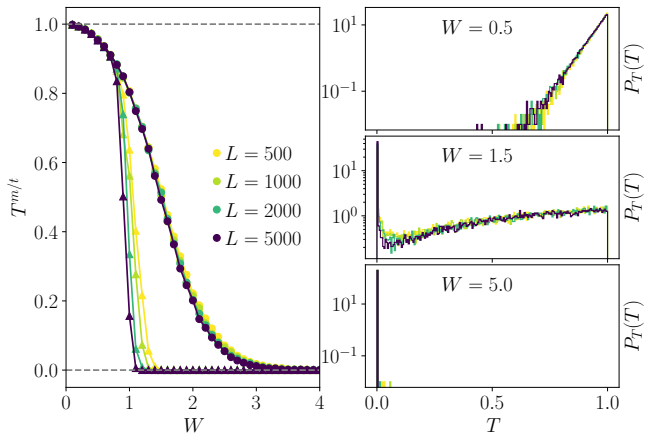

FIG. S1. Left: The mean (circles) and typical (triangles) transmittance, $T^{m/t}$, as a function of $W$, for various system sizes $L$. Data are computed over 5000 realisations. Right: Distributions of transmittances $P_T(T)$ for the same system sizes, for $W=0.5$, $1.5$ and $5$, in the delocalised, mixed and multifractal regimes respectively. 20000 disorder realisations were used to form each distribution.

The disorder average of Eq. S4, $\langle \mu_E^2 \rangle = W^2(1 - \langle S_1^2 \rangle) + 2$, is the variance of the disorder-averaged spectrum. However, $\mu_E^2$ also fluctuates from realisation-to-realisation. Computing $\langle \mu_E^4 \rangle$, it is readily shown that

$$\mathrm{var}(\mu_E^2) = \langle \mu_E^4 \rangle - \langle \mu_E^2 \rangle^2 \sim \mathcal{O}(W^4). \quad (S5)$$

The second moment of the density of states therefore also fluctuates between realisations, and importantly does not decrease with increasing system size. These strong fluctuations explain the presence of the mixed region of extended and multifractal states: for a given energy, different disorder realisations host an extended-multifractal eigenstate phase transition at different values of $W$, leading to a finite region of $W$ in which some realisations are in the extended phase and some in the multifractal phase.



\end{document}